\documentclass[aps,pra,showpacs,amssymb,nofootinbib,superscriptaddress,twocolumn,longbibliography]{revtex4-1}
\usepackage[T1]{fontenc}
\usepackage[english]{babel}
\usepackage[utf8]{inputenc}
\usepackage{amsmath,amssymb}
\usepackage[colorlinks=true,citecolor=blue]{hyperref}
\usepackage{graphicx}
\usepackage{lipsum}% http://ctan.org/pkg/lipsum
\usepackage{amsmath}
\usepackage{slashed}
\usepackage[table,xcdraw,dvipsnames]{xcolor}
\usepackage{bbm}
\usepackage{soul}
\usepackage{array,booktabs}
\usepackage{natbib}
\usepackage{amsmath}
\usepackage{bm}% bold math
\usepackage{bbold}
\usepackage{hhline}
\usepackage{xcolor}
\usepackage{mathtools} % dcases
\usepackage{float}
\usepackage{tabularx}
\usepackage{tikz}
\usepackage{physics}
\usepackage{MnSymbol}

\newcommand{\adag}{\hat{a}^{\dagger}}
\newcommand{\bdag}{\hat{b}^{\dagger}}
\newcommand{\LOOL}{\ell 00 \ell}

\begin{document}

\title{$\LOOL$ entanglement and the twisted quantum eraser}

\author{Dylan Danese}
\email{dd52@hw.ac.uk}
\affiliation{Institute of Photonics and Quantum Sciences, Heriot-Watt University, Edinburgh EH14 4AS, UK}

\author{Sabine Wollmann}
\affiliation{Institute of Photonics and Quantum Sciences, Heriot-Watt University, Edinburgh EH14 4AS, UK}

\author{Saroch Leedumrongwatthanakun}
\thanks{Current address: Division of Physical Science, Faculty of Science, Prince of Songkla University, Songkhla 90110, Thailand}
\affiliation{Institute of Photonics and Quantum Sciences, Heriot-Watt University, Edinburgh EH14 4AS, UK}

\author{Will McCutcheon}
\affiliation{Institute of Photonics and Quantum Sciences, Heriot-Watt University, Edinburgh EH14 4AS, UK}

\author{Manuel Erhard}
\affiliation{Quantum Technology Laboratories, Clemens-Holzmeister-Straße 6/6, 1100 Vienna, Austria}

\author{William N. Plick}
\affiliation{33 Charles St. E. M4Y 0A2, Toronto, ON, Canada}

\author{Mehul Malik}
\email{m.malik@hw.ac.uk}
\affiliation{Institute of Photonics and Quantum Sciences, Heriot-Watt University, Edinburgh EH14 4AS, UK}

\begin{abstract}
We demonstrate the generation of unbalanced two-photon entanglement in the Laguerre-Gaussian (LG) transverse-spatial degree-of-freedom, where one photon carries a fundamental (Gauss) mode and the other a higher-order LG mode with a non-zero azimuthal ($\ell$) or radial ($p$) component. Taking a cue from the $N00N$ state nomenclature, we call these types of states $\LOOL$-entangled. They are generated by shifting one photon in the LG mode space and combining it with a second (initially uncorrelated) photon at a beamsplitter, followed by coincidence detection. In order to verify two-photon coherence, we demonstrate a two-photon ``twisted'' quantum eraser, where Hong-Ou-Mandel interference is recovered between two distinguishable photons by projecting them into a rotated LG superposition basis. Using an entanglement witness, we find that our generated states have fidelities of 95.31\% and 89.80\% to their respective ideal maximally entangled states. Besides being of fundamental interest, this type of entanglement will likely have a significant impact on tickling the average quantum physicist's funny bone.

%We demonstrate the experimental generation and verification of a "L00L" state using a spatial light modulator and full state tomography with a fidelity of $0.94$.
\end{abstract}

% \begin{document}

\maketitle

%%%%%%%%%%%%%%%%%%%%%%%%%%  body  %%%%%%%%%%%%%%%%%%%%%%%%%%
\section{introduction}

Jon Dowling was a methodical and effective researcher, but when it came to \emph{talking} about research \--- either in a plenary talk or at the coffee shop \--- the man liked to ham it up. He took every opportunity to insert humor and wackiness wherever he could find a crevice to jam it in, even if it didn't \emph{really} belong. It was rare to see him give a talk on $N00N$ states without a poster from Gary Cooper's ``High Noon'' and peppering his explanations with gun-slinger analogies \--- even if this was at a government program review. Part of the audience would roll their eyes and groan, some would snicker, while the rest (as is customary) would be zoned out working on their own talks or sleeping off the night before.

From the perspective of a young impressionable graduate student, facing down rows of dour professors, Jon was both a relief and a revelation. A relief from the self-importance and humorlessness of much of the community, and proof that science was \emph{allowed to be fun too}, and that just because something was silly didn't mean that it wasn't important. 

Jon loved science fiction and fantasy as well, and when students would be preparing slides to present to Department of Defense officials he would claim that they could be mollified by inserting some ``Borg soldiers with laser beams running around!'' in locations through slides proximate to phrases like ``annihilation operator'' (in addition to stealth bombers and reconnaissance satellites placed in the background). Since the topic of this article is the most recent of a long chain of elaborations, generalizations, and extensions of Jon's beloved $N00N$ states, it seems fitting to tell a story that will (eventually) fit the theme.

WNP, one of Jon's early PhD students (and co-author here), had attended the 2018 QCMC Conference as a newly-minted assistant professor himself. Jon was of course quite busy, having organized the whole thing, and WNP had not had the chance to spend that much time with him socially. At the conference banquet (held at a stately plantation deep within the Louisiana Bayou) WNP was waiting in line to get a drink at the cash bar. ``Bildo!'' WNP heard Jon exclaim unexpectedly from behind (this was Jon's affectionate nickname for him). Jon quickly joined him in line (ahead of a number of others) and took the opportunity to catch up, chat, and reminisce. 

\begin{figure}
\includegraphics[width=0.45\textwidth]{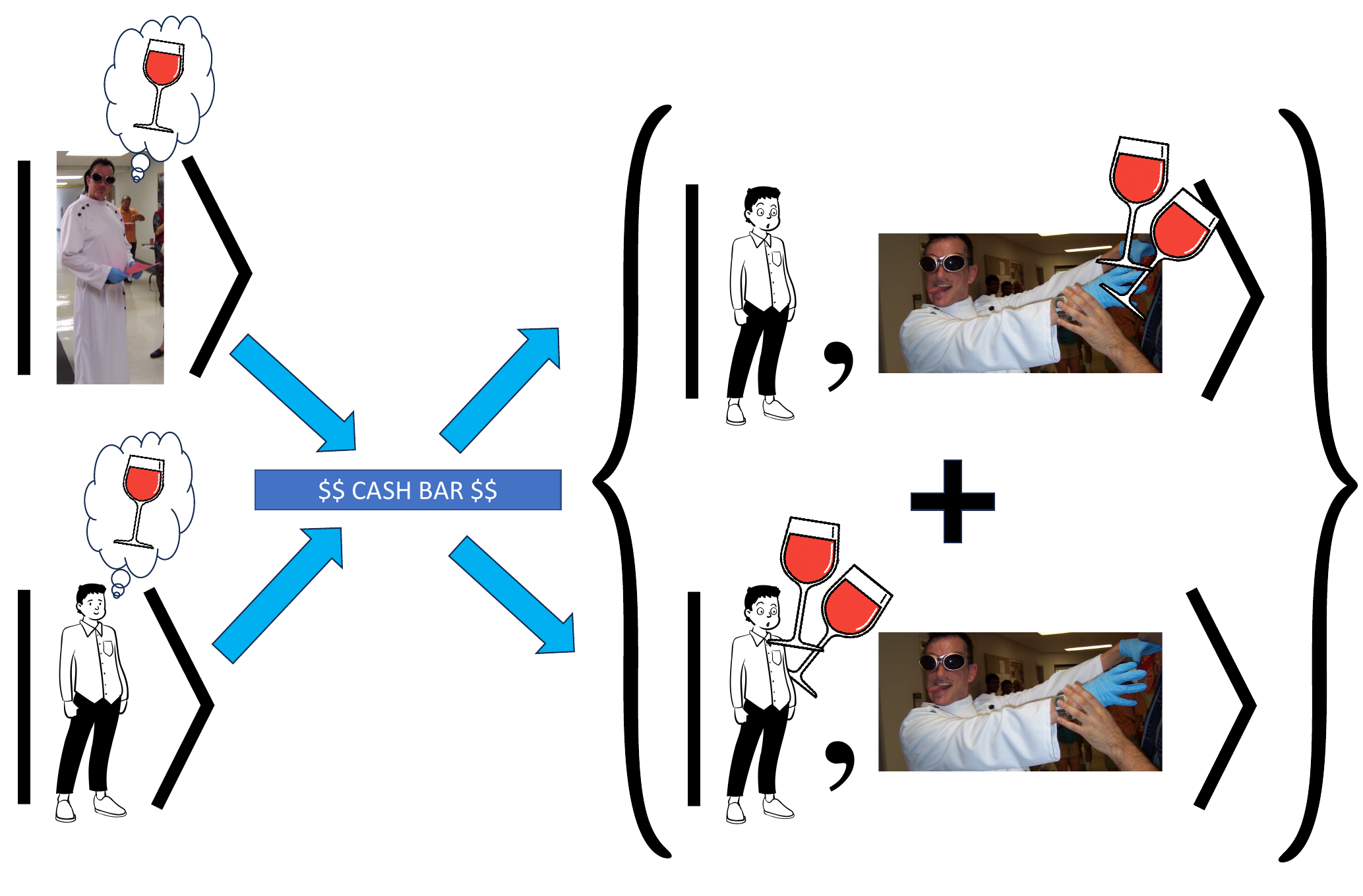}
\caption{Jon Dowling demonstrating the not-so-quantum mechanical phenomenon of wine-bunching at QCMC 2018. \label{fig:WB}}
\end{figure}

``Bill, you got a tab open?'' Jon asked when they arrived at the front, making a show of patting his (supposedly) empty pockets. WNP confirmed that he did, and Jon asked if he minded putting his wine on it \--- WNP, no longer technically poor, agreed happily. When the wine came Jon grabbed ahold of \emph{both} glasses, exclaimed ``Thanks!'' loudly and hurried off laughing. 

Was Jon just stealing wine from his former student? Or was he once again demonstrating photon bunching? Two similar particles \--- thirsty theoreticians who drink maybe too much and have a questionable sense of humor \--- are incident on a cash bar. Though both modes continue before and after the interaction, the wine (mode occupation) tends to go in one direction or the other (see Fig.~\ref{fig:WB}). Perhaps taking the many-worlds view of quantum mechanics, there is another universe where the superposition was realized in the other direction and WNP ended up with both of the wines that he payed for, instead of none. WNP likes to think that in that other universe Jon is still around as well, to continue to inspire people to have fun and do great science.    

First discussed in the context of quantum decoherence in 1989 \cite{Sanders1989}, the idea of a $N00N$ state was, to quote Jon, rediscovered by Jon's group in 2000 for applications in quantum imaging, in particular quantum lithography \cite{boto2000quantum}. As their moniker implies, $N00N$ states describe a two-mode entangled state consisting of a coherent superposition of modes A and B carrying photon numbers $\ket{N}_\textrm{A}\ket{0}_\textrm{B}$ or $\ket{0}_\textrm{A}\ket{N}_\textrm{B}$. Such states can be experimentally realised via linear optical circuits involving interference between one or more photons propagating in multiple modes \cite{cable2007efficient, shin2013enhancing}. The close relationship between ``digital'' or discrete-variable photonic quantum metrology and quantum computing was astutely pointed out by Jon and coworkers \cite{Lee2002,dowling2008quantum}, where he predicted that realistic applications of the former are likely to precede the latter. The past two decades have seen extensive debate about the quantum metrological advantage of $N00N$ states, especially when all experimental resources are accounted for \cite{Resch2007, Slussarenko2017}. A parallel line of research, motivated by progress in our ability to structure quantum light \cite{Malik2014}, has ventured in the direction of ``twisted'' $N00N$ states \cite{Jha2011, Dambrosio2013, Hiekkamaki2021, hiekkamaki2022observation}. The idea here is to combine the Heisenberg-limited phase supersensitivity of $N00N$ states with shot-noise-limited angular superresolution obtained by imprinting the photons with orbital angular momentum (OAM), resulting in an increased sensitivity to rotation.

Here, we build on the idea of a $N00N$ state to explore the generation of a spatially entangled state where the photon number $N$ is replaced by the spatial mode index, and the photon number in each mode is fixed to be one. Working in the Laguerre-Gaussian (LG) tranverse spatial mode basis \cite{Krenn2017Orbital}, where modes are labelled by their azimuthal ($\ell$) or radial ($p$) indices, we get the pleasing terminology of $\LOOL$ states\footnote{In order to spare posterity the indignity, here we have refrained from the obvious extension of $\LOOL$ states to the radial index $p$}. Here, the photons are in a coherent superposition of carrying a fundamental Gaussian mode ($\ell=p=0$) and a higher order mode ($\abs{\ell}>0$, $p>0$). As the LG mode space is unbounded, these states can also be classified as high-$\LOOL$ states, making them candidates for the hotly debated Schr\"{o}dinger-cat states \cite{Schrodinger1935}. The question of macroscopicity is raised when one considers that very large amounts of angular momentum could conceivably be transferred to quantum states of matter, as has been demonstrated (for low OAM quanta) with Bose-Einstein condensates and clouds of cold Rubidium atoms \cite{Andersen2006, Parigi2015}. Interestingly, $\LOOL$ states have been seen before as a key building block in computer-designed experiments on high-dimensional multi-photon entanglement \cite{Melvin, 332, OAMGHZ}. The generation of the first three-particle, three-dimensional GHZ state required the generation of an $\LOOL$ state with $\ell=2$ as an intermediate step, and appears in several other experimental designs found via a computational algorithm. Below, we describe in detail how $\LOOL$ states can be created using linear optical elements followed by post-selection on two-photon coincidences, and demonstrate their experimental generation and characterisation via an entanglement witness and quantum state tomography.

\section{Theory} \label{sec:theory}

Here we describe how an $\LOOL$-entangled state can be generated from two initially uncorrelated single photons. For conciseness, we will limit our discussion to $\LOOL$ states, but the same theory applies to the radial index $p$. Consider an input state consisting of two single photons carrying arbitrary values of orbital angular momentum or OAM ($\ell_1$ and $\ell_2$) in orthogonal paths ($a$ and $b$):

\begin{equation} \label{eq:Psi_in}
    \ket{\psi_{\text {in }}} = \adag_{\ell_1} \bdag_{\ell_2}\ket{\mathrm{vac}} = \ket{\ell_1,\ell_2}_{a b}
\end{equation}

\noindent Note that in contrast with the notation used in the $N00N$ state literature, the numbers inside the kets here refer to the OAM mode carried by a single photon and \underline{\textbf{not}} the photon number. An implicit assumption is made that a single photon ($N=1$) is present in each mode and that higher photon numbers are absent. We also assume that the two photons are perfectly indistinguishable in all other degrees of freedom (spectrum, polarization, etc). These two photons are then incident on two input ports of a 50:50 beam splitter (BS). This state is unitarily transformed by the BS as follows.
\begin{figure}[t]
\centering
\includegraphics[width=0.47\textwidth]{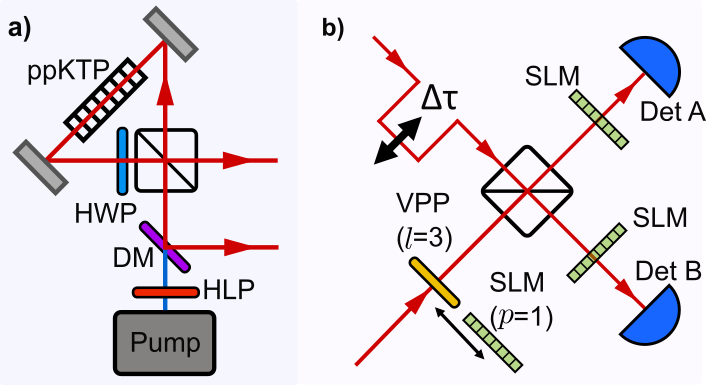}
\caption{(a) Two-photon source: A pump laser at 775nm is used to generate a pair of photons at 1550nm via SPDC in a nonlinear ppKTP crystal placed inside a Sagnac loop. A dichroic mirror (DM) filters the pump from the generated photon pairs. As the pump is horizontally polarized (HLP), the ppKTP is pumped only in one direction. The generated photons are coupled into single mode fibers (SMF). (b) $\LOOL$ entanglement generation: The two fiber-coupled photons are launched towards a $50:50$ beamsplitter (BS). A translation stage ($\Delta\tau$) introduces a variable path delay for one photon, while the other passes through a vortex phase plate (VPP) that adds an OAM value of $\ell=3$. Alternatively, a spatial light modulator (SLM) is used to imprint an LG mode with $p=1$ onto the photon. The combination of an SLM, SMF, and a superconducting nanowire single-photon detector (SNSPD, not shown) are used for performing generalized LG projective measurements on each output photon (A and B) arriving from the BS. The single-photon detection events are correlated using a coincidence counting logic (not shown).}
\label{fig:setup}
\end{figure}

\begin{equation} \label{eq:state_BS}
    \ket{\psi_{\text{out}}} = \frac{1}{2}\left(\adag_{\ell_1} \adag_{\ell_2} + \adag_{\ell_1} \bdag_{\ell_2} -\adag_{\ell_2} \bdag_{\ell_1} - \bdag_{\ell_1} \bdag_{\ell_2}\right)\ket{\mathrm{vac}}
\end{equation}

\noindent As seen in eq.\ref{eq:state_BS}, there are four terms arising from the reflection or transmission of each photon on the BS. By virtue of coincidence detection, we select only the events that result in one photon being present in each output port. Consequently, the post-selected two-photon state can be written as the maximally entangled state:

\begin{equation}
    \ket{\chi}=\frac{1}{\sqrt{2}}\Big(\ket{\ell_1,\ell_2}-\ket{\ell_2,\ell_1}\Big).
\end{equation}

\noindent It is important to note that reflection at the BS flips the sign of OAM modes $\ell_1$ and $\ell_2$, which is remedied by the inclusion of an extra reflection in input path $a$ and output path $b$. Note also that while we have renormalized the state here, it is obtained after the BS with a probability of one-half.  This is a Bell state where the photons are supported on the modes, $\ket{\ell1}$ and $\ket{\ell2}$. In contrast to a high-dimensional OAM-entangled two-photon state directly generated through a nonlinear process such as spontaneous parametric downconversion \cite{LGEnt}, this state does not result from the conservation of pump momentum. Instead, it is the result of a coherent superposition of the two output possibilities from the BS. As such, the OAM values involved can be arbitrarily chosen at the input stage. By setting the OAM value carried by the second input photon to zero ($\ell_2=0$) we can then realize the $\LOOL$ state:

\begin{equation}\label{eqn:l00l_state}
\ket{\chi} = \frac{1}{\sqrt{2}}\Big(\ket{{\ell},0}-\ket{{0,\ell}}\Big).
\end{equation}

\noindent Let us investigate the coherence properties of this entangled state further. For instance, while the state shows correlations in the computational basis $\{\ket{0},\ket{\ell}\}$, it should also exhibit correlations in the superposition basis $\ket{\pm} = (\ket{0} \pm \ket{\ell})/\sqrt{2}$. While theoretically this is quite straightforward, experimentally it results in interesting two-photon interference effects. Two-photon or Hong-Ou-Mandel (HOM) interference is normally observed between two indistinguishable photons incident on a 50:50 BS---the probability amplitude corresponding to one photon being transmitted and the other reflected destructively interferes with the opposite case and both photons are either transmitted or reflected, resulting in a so-called HOM dip \cite{Hong1987}. This behaviour is also observed with Bell states, where the three states $\ket{\phi_+}$, $\ket{\phi_-}$, and $\ket{\psi_+}$ exhibit a HOM dip (bunching), and the $\ket{\psi_-}$ exhibits a HOM bump (antibunching).

Since the two photons incident on the BS above carry different values of OAM, they are in orthogonal modes with respect to each other, making them distinguishable and unable to exhibit quantum interference. However, by measuring in the superposition basis, one can erase this distinguishing information and recover two-photon quantum interference. This is akin to the standard quantum eraser \cite{Scully1982}, where single-photon interference is recovered by erasing the distinguishing information between the two modes of an interferometer by projecting the state into a superposition of the two. A two-photon quantum eraser was first demonstrated with the polarisation degree-of-freedom in 1992 \cite{Kwiat1992}. Here we examine our ``twisted'' two-photon case further by considering the OAM superposition projection operators $\ket{+}\bra{+}$ and $\ket{-}\bra{-}$. We construct two measurements on photons A and B, $\mathcal{P}_{\text{sym}} = \frac{1}{4}\ket{+}\bra{+}_A \otimes \ket{+}\bra{+}_B$ and $\mathcal{P}_{\text{asym}} = \frac{1}{4}\ket{+}\bra{+}_A \otimes \ket{-}\bra{-}_B$, which we call the symmetric and antisymmetric projectors respectively. They are proportional (up to postselection) to projections on the Bell states $\ket{ \Phi^+}$ and $\ket{\Psi^-}$ before the beamsplitter, which are symmetric and antisymmetric (with respect to path exchange) respectively. The results of performing these two sets of measurements are as follows:

\begin{figure*}[t!]
\centering
\includegraphics[width=0.9\textwidth]{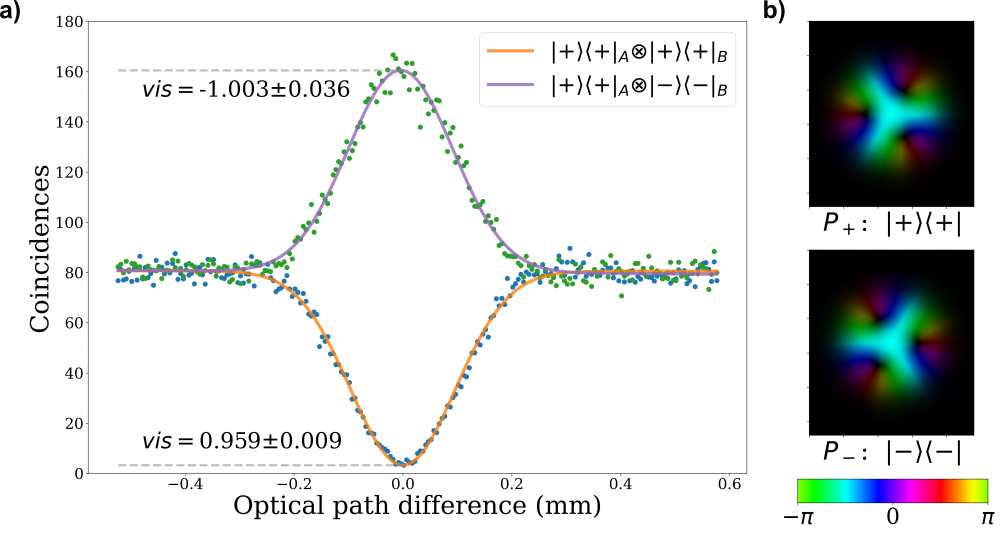}
\caption{ A two-photon ``twisted'' quantum eraser: Bunching and anti-bunching observed between two initially distinguishable photons by performing suitable projective measurements on them. The blue dots fitted with the orange curve represent measurements with the symmetric holograms (when Bob and Alice both measure the same superposition state $\ketbra{+}{+}$).  We get a HOM dip with 0.959 visibility (blue dots fitted with orange curve). Conversely,  The green dots fitted with the purple curve represent measurement with the anti-symmetric holograms (Alice measures $\ketbra{+}{+}$ while Bob measures $\ketbra{-}{-}$), where we get a bump with a visibility of -1. In both cases we measure the raw coincidences. Note that $ \ket{\pm} =\frac{1}{\sqrt{2}}(\ket{0}\pm\ket{3})$.}
\label{fig:HOM_results}
\end{figure*}

\medskip

\begin{equation}\label{eqn:expectation_value_l00l}
    \begin{aligned} 
        \bra{\chi}\mathcal{P}_{\text{sym}}\ket{\chi} &= 0 \\
        \bra{\chi}\mathcal{P}_{\text{asym}}\ket{\chi} &= \frac{1}{2}.
    \end{aligned} 
\end{equation}

The symmetric projection operating on the $\ell00\ell$ state results in complete destructive two-photon interference in post-selection, i.e., photon-bunching due to the erasure of distinguishing information between the two photons. While the antisymmetric operator also performs information erasure, it however results in constructive two-photon interference, i.e., photon anti-bunching. This follows from the $\LOOL$ state being anti-symmetric from path exchange. In contrast, let us perform these measurements on a classically correlated mixed state in the $\{\ket{0},\ket{\ell}\}$ subspace, $\rho_{c} = \frac{1}{2}\left(\ketbra{\ell0}{\ell0} + \ketbra{0\ell}{0\ell}\right)$. In this case, the measurements above result in 
\begin{equation}\label{eqn:expectation_value_mixed}
    \Tr{\rho_{c}\mathcal{P}_{\text{sym}}} = \Tr{\rho_{c}\mathcal{P}_{\text{asym}}} = \frac{1}{4}.
\end{equation}

\noindent As can be seen, the classically correlated state gives an intermediate value for both symmetric and antisymmetric projectors due to the lack of coherence. This can be seen as the transition from HOM dip/bump to no interference, as two-photon coherence is lost when the photons are made distinguishable, for example due to different arrival times.

\section{Experiment}

A simplified schematic of the experimental setup can be seen in Fig.~\ref{fig:setup}. We generate photon pairs at 1550nm via the process of Type-II spontaneous parametric downconversion (SPDC) in a ppKTP crystal pumped by a narrowband laser at 775nm. The crystal is placed in a Sagnac configuration normally used for the generation of polarization entanglement at telecom wavelengths via bidirectional pumping \cite{Proietti2021}. However, since here we only want to generate separable photons, the crystal is pumped only in one direction by setting the pump polarization to be horizontal. The generated signal and idler photons are then coupled into single-mode fibers (SMF) and transported to the entanglement generation setup where they are launched into free-space.  One photon passes through a trombone system of mirrors placed on a motorized translation stage which introduces a variable path delay. The second photon passes through a vortex phase plate (VPP) that imprints a transverse phase structure onto it corresponding to an orbital angular momentum (OAM) of $\ell=3$. The photons are then interfered on a 50:50 beamsplitter (BS) and the outgoing paths are routed to spatial light modulators (SLM, Holoeye Pluto-2.1 LCOS). A blazed diffraction grating with a suitable period is displayed on the SLMs and the first diffraction orders are coupled into single-mode fibres leading to superconducting nanowire single-photon detectors (SNSPDs). Correlated detection events within a time window of 0.2 ns are recorded with a coincidence counting logic (Time Tagger Ultra, Swabian instruments). The combination of an SLM, SMF, and SNSPD acts as a programmable generalised projective measurement \cite{bouchard2018measuring}, allowing us to make measurements of LG modes and their coherent superpositions. In order to ensure the performed measurements correspond to those defined on the exit ports of the beam splitter, the effects of mirrors (taking $\ket{ \ell}\rightarrow \ket{-\ell}$) and lenses altering the beam waists on the SLM planes must be taken into account.  

\medskip

In order to ensure temporal indistinguishability between the two photons, a motorized ``trombone'' system of mirrors is used to introduce a variable path length delay ($\Delta\tau$) between the two photons. We initially measure the indistinguishability of the two Gaussian mode photons without using the vortex phase plate (VPP) in Fig.~\ref{fig:setup}, where the SLMs only display a holographic diffraction grating in order to couple the photons directly to the SMFs. We obtain a HOM dip with a visibility of $0.96$, indicating that the photons are highly indistinguishable. Here we define visibility as $\mathrm{Vis} = \frac{C(\delta = \inf) - C(\delta = 0)}{C(\delta=\inf)}$, where $C$ is our fitted Gaussian curve \cite{ou1999photon}. This definition yields $1$ for perfect photon bunching and $-1$ for perfect anti-bunching.

\medskip

Next, an $\ell=3$ vortex phase plate is placed in one of the input paths. Now, to observe two-photon interference again, we erase the distinguishing information between the two photons (Section~\ref{sec:theory}) by performing the projection operators corresponding to different LG mode superpositions. In practice, this is carried out by using the SLM and SMF together as a projective measurement \cite{valencia2020high,bouchard2018measuring}. For instance, if we want to measure the equal superposition mode, $ \ket{+} =\frac{1}{\sqrt{2}}(\ket{0}+\ket{3})$, we create a computer generated hologram (CGH) \cite{arrizon2005accurate} corresponding to the complex conjugated field of this superposition mode and then display it on the SLM. In this manner, we can experimentally realize the projective measurements outlined in section~\ref{sec:theory}. The same technique is used to perform local projective measurements in all mutually unbiased bases (MUBs) of the two-dimensional state space and record coincidence counts. This data constitutes a tomographically complete set of measurements which can be used to reconstruct the entire density matrix of the $\LOOL$ state via full quantum state tomography. We employ a linear inversion with projected least squares for a practical reconstruction of the state ~\cite{guctua2020fast}.

\begin{figure}[t!]
\centering
\includegraphics[width=0.45\textwidth]{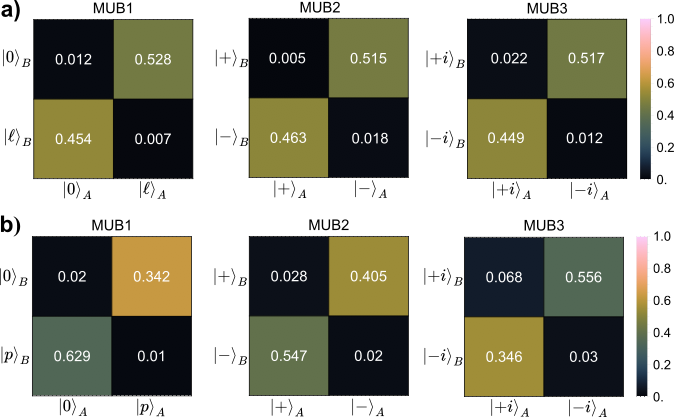}
\caption{Measured coincidence counts (normalised) in the three mutually unbiased bases (MUBs) required for evaluating the entanglement witness.  Two-photon correlations are strongly preserved across the MUBs, where a) a $\LOOL$ state (with $\ell=3$) and b) the radial equivalent (with $p=1$) achieve fidelities of $F=0.9531\pm0.0061$ and $F=0.8980\pm0.0053$ to their respective maximally entangled states. A and B subscripts denote measurements made at Alice or Bob, and MUB 1 is the computational basis, where $\ket{0}$ denotes the Gaussian mode and $\ket{\ell/p}$ denotes a mode with non-zero LG index ($\ell/p$). MUBs 2 and 3 are superposition bases where, for a) $ \ket{\pm} =\frac{1}{\sqrt{2}}(\ket{0}\pm\ket{\ell})$ and $ \ket{\pm i} =\frac{1}{\sqrt{2}}(\ket{0}\pm i\ket{\ell})$. In b), the LG mode index $\ell$ is replaced by the radial mode $p$ for all MUBs. \label{fig:MUBWitnessData}}
\end{figure}

\medskip

In order to create a radial mode entangled state, we next replace the vortex phase plate shown in Fig.~\ref{fig:setup}b with an SLM displaying a computer-generated hologram for a radial LG mode with index $p=1$. Analogous to the previous case, the two-photon state is now entangled across the fundamental Gaussian mode and the $p=1$ mode by simply replacing the $\ell$ with $p$ in Eq.~\eqref{eqn:l00l_state}. Since we are using an SLM to generate this state, the experimental setup needs to be modified to allow for the SLM to be used in reflection rather than transmission as for the VPP (not shown in simplified setup in Fig.~\ref{fig:setup}). Once the radial mode hologram is implemented, two-photon coherence is found again by searching for a HOM dip. Again, we employ computer-generated holography to create phase masks corresponding to superposition measurements in the $p=0$ and $p=1$ subspace. Finally, we perform projective measurements in all MUBs of this two-dimensional subspace and reconstruct the state using quantum state tomography in the same manner as above.

% \begin{equation}\label{eqn:poop_state}
% \ket{\chi} = \frac{1}{\sqrt{2}}\Big(\ket{{p},0}-\ket{{0,p}}\Big).
% \end{equation}

\section{Results and Discussion} 

\begin{figure}[b!]
\centering
\includegraphics[width=0.45\textwidth]{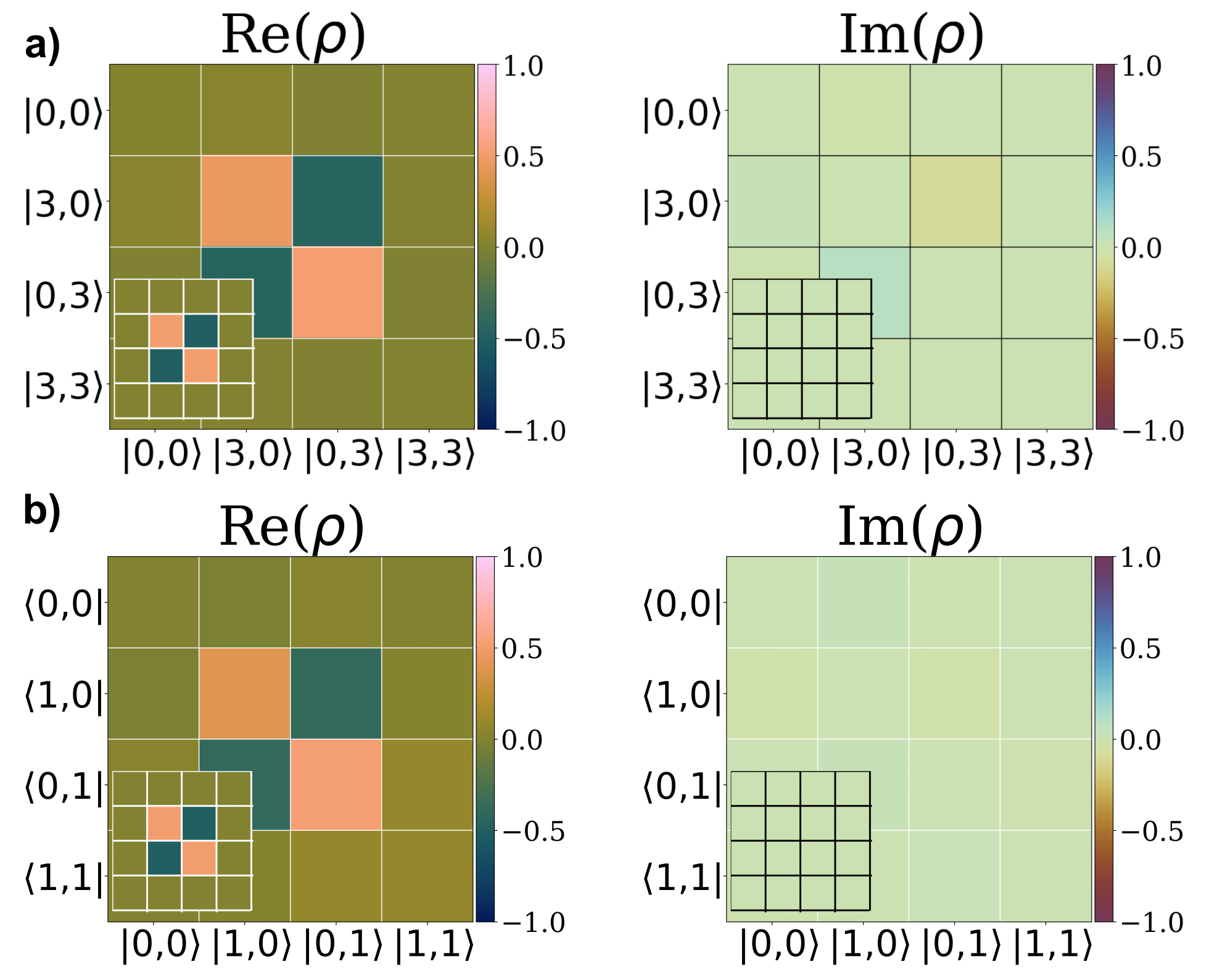}
\caption{Density matrices of the recovered a) $\LOOL$ state with $\ell=3$ and b) radial mode counterpart with $p=1$, alongside the ideal expected states (insets). \label{fig:density_matrix}}
\end{figure}

Two-photon interference measured by implementing the sets of projection operators $\mathcal{P}_{\text{sym}}$ and $\mathcal{P}_{\text{asym}}$ on SLMs A and B are shown in Figure~\ref{fig:HOM_results}a. The HOM dip and bump are observed with visibility of $0.92$ and $1$ respectively, indicating the presence of two-photon coherence in the $\LOOL$ state. This is in strong agreement with the theory discussed in section~\ref{sec:theory}. Figure~\ref{fig:HOM_results}b shows computer-generated holograms that are used for measuring the superposition states $ \ket{\pm} =\frac{1}{\sqrt{2}}(\ket{0}\pm\ket{3})$. In practice, however, these holograms are implemented with a blazed diffraction grating to remove noise from zero-order diffraction. Next, we use an entanglement witness based on measurements in mutually unbiased bases (MUBs) \cite{Bavaresco2018} to calculate the state fidelity $F = \bra{\chi} \rho \ket{\chi}$  with respect to the ideal $\LOOL$ and radial mode entangled states (Eq.~\eqref{eqn:l00l_state}). As shown in Fig.~\ref{fig:MUBWitnessData}, strong correlations are observed in all three MUBs, resulting in a fidelity of F=$0.9531\pm0.0061$ and F=$0.8980\pm0.0053$ for the $\LOOL$ and radial mode entangled states, respectively. Finally, we reconstruct the state density matrices via quantum state tomography. The real and imaginary parts of both density matrices are shown in Fig.~\ref{fig:density_matrix}, and demonstrate good agreement with the theoretically expected values (insets). 

The reduction in HOM visibility when inserting the vortex plate, and the small amount of deviation from the ideal state fidelity is likely due to the non-ideal nature of the generated $\ell=3$ mode. The VPP imprints an $\ell=3$ spiral phase on the input Gaussian mode, which is known to approximate a Laguerre-Gaussian mode upon propagation \cite{Vallone2016}. However, it has a non-ideal overlap with an LG mode with $\ell=3$, which results in crosstalk and a consequent reduction in HOM visibility and state fidelity. In addition, a slight misalignment of the vortex plate from the optical beam axis and the non-ideal nature of SLM projective measurements \cite{bouchard2018measuring, Qassim2014} might further contribute to a reduction in the state quality. However, as shown recently, the effects of such measurement imperfections can be calibrated and corrected for \cite{Goel2022}. Finally, one should note that this scheme can at most generate a two-dimensional entangled state, and does not trivially generalise to producing genuine high-dimensional entangled states \cite{LGEnt,PixelEnt}. However, schemes based on path-identity have been found and demonstrated that can produce high-dimensional entanglement from initially uncorrelated photons, which also incorporate a mode-shift in the manner shown here \cite{Krenn2017Path, Kysela2020}. 

In conclusion, we have demonstrated the generation of two-photon entanglement in the Laguerre-Gaussian (LG) degree-of-freedom from two initially uncorrelated photons. The entanglement takes on an unbalanced form, with one photon carrying a fundamental Gaussian mode and the other a higher-order azimuthal or radial LG mode. Taking inspiration from the $N00N$ state nomenclature, the resulting entangled states can be written as $\LOOL$. The states are generated by shifting one photon in the LG mode space and coherently recombining it with another photon at a beamsplitter, and post-selecting on two-photon detection events. Besides being of fundamental and humorous interest, such generation schemes can be seen as being a building block in methods for generating complex forms of entanglement \cite{OAMGHZ,Krenn2017Path} and could lead to interesting research directions using other photonic degrees-of-freedom, such as frequency ($\omega00\omega$ states) or time ($t00t$ states).

\begin{acknowledgments}
We would like to express our gratitude to Suraj Goel, Dr. Natalia Herrera Valencia and Vatshal Srivastav for lending their expertise and valuable time to discuss the implementation of this project. This work was made possible by financial support from the UK Engineering and Physical Sciences Research Council (EPSRC) (EP/P024114/1), the European Research Council (ERC) Starting grant PIQUaNT (950402), and the Royal Academy of Engineering under the Chair in Emerging Technologies Programme (CiET-2223-112). SW acknowledges funding from the European Union’s Horizon 2020 research and innovation programme under the Marie Skłodowska-Curie grant agreement No 89224.
\end{acknowledgments}

\bibliography{LOOLbib}

\end{document}